\documentclass{aa}
\usepackage{graphicx}
\begin{document}
%
   \title{An upper bound on the energy of a gravitationally redshifted  electron-positron
annihilation line from the Crab pulsar
    }

   \author{P. S. Negi
          }


   \institute{Department of Physics, Kumaun University,
              Nainital - 263 002, India\\
              \email{psnegi\_nainital@yahoo.com; negi@aries.ernet.in}
             }
\maketitle
A\&A 431, 673-677(2005); DOI: 10.1051/0004-6361:20041790

\keywords{dense matter - equation of state - stars:
                neutron - stars: pulsars: individual: Crab
               }

 
I have found an error in the numerical calculations of the envelope component of the model discussed in my paper A\&A 431, 673-677 (2005). The removal of this error and the extension of calculations for the whole model show minute changes in some of the results whereas rest of them vary significantly from those published in the paper (hence also the Figs.(1)-(3) and Table 1 are not correct). The corrected results are summarized pointwise (from point (1) to (5)) below, indicating page number, paragraph or line number of the published paper, right from the begining (abstract) to the end (results and conclusions). In order to utilize the space limitation of the erratum, the results of detailed calculations for the model are also summarized in Table 1. It follows from Table 1 that this model is also capable of explaining the glitch healing parameters of Vela pulsar together with the Crab. But this issue is out of the scope of this erratum and will be addressed in a forthcoming paper (arXiv: 0710.4442).

(1) The third sentence of `abstract' should be replaced by the following sentence: The models yield an upper bound on surface redshift, $z_R \simeq 0.77$, of 
neutron stars corresponding to the case of $\Gamma_1 = 2$ envelope, whereas a model-independent
upper bound on neutron star masses, $M_{\rm max} \leq 4.2 M_\odot$, is obtained for a conservative
choice of the `matching density', $E_b = 2.7 \times 10^{14}{\rm\,g\,cm}^{-3}$, at the core-envelope
boundary. Rest of the statement of `abstract' right from the fourth sentence should be replaced by the following statement: If the lower limit of the observational constraints of (i) the glitch healing parameter, $Q \geq 0.7$, of the Crab pulsar and (ii) the recently evaluated 
value of the moment of inertia for the Crab pulsar, $I_{\rm Crab,45} \geq 3.04$ (where $I_{45}=I/10^{45}\,{\rm g.cm}^2$), both are  imposed together on 
these models, the model with a  $\Gamma_1 = 2$ envelope yields the value of matching density, $E_b = 7.0794 \times 10^{14}{\rm\,g\,cm}^{-3}$. This value of matching density yields a model-independent upper bound on neutron star masses, $M_{\rm max} \leq 2.6 M_\odot$, and the strong lower bounds on surface redshift $z_R \geq 0.6232$ and mass $M \simeq 2.455 M_\odot$ for the Crab pulsar. However, for the observational constraint of the `central' weighted mean value $Q \approx
0.72$, and $I_{\rm Crab,45} > 3.04$, the minimum surface redshift and mass of the Crab pulsar are slightly increased to the values $z_R \simeq 0.6550$ and  $M \simeq 2.500 M_\odot$ respectively. The confirmation of these results requires evidence of the observation of the 
gravitationally redshifted electron-positron annihilation line in the energy range of about 0.309 - 0.315 MeV from the Crab pulsar.

(2) Page 675, left column, first line, ``if the {\em minimum} value...as shown in Table 1'' of section 2 should be replaced by the following statement: if the {\em minimum} value
 of the ratio of pressure to energy-density, $P_b/E_b$, at the core-envelope boundary reaches 
about $4.694 \times 10^{-2}$. The results of the study are presented in Table 1 for a conservative choice 
of $E_b = 2.7 \times 10^{14}\,{\rm g\, cm}^{-3}$, the nuclear saturation density. It is seen that the models 
become pulsationally stable up to the maximum value of mass $M_{\rm max} \approx 4.2 M_\odot$ and radius, 
$R \simeq 
18.16 - 20.37 {\rm \,km}$. The minimum radius results for the model with a $\Gamma_1 = 2$ envelope thus 
maximizes the compactness ratio for the stable configuration, $u \simeq 0.34$, as shown in Table 1.
The second last sentence of section 2 on page 675 is removed and in the third line of the last sentence of this section, the value 0.3235 is replaced by 0.3201. Furthermore, the following sentence is to be added at the end of section 2: However, for  $\Gamma_1 = (4/3)$ envelope model the binding energy reaches a maximum beyond the maximum value of mass.

(3) Page 675, right column, last paragraph of section 3 ``Equation (6) is used...surface redshift of NSs.'' is to be replaced by the following paragraph: 
Equation (6) is used, together with coupled Eqs.(3) - (5), to calculate the fractional moment of inertia
given by Eq.(1) and the moment of inertia of the entire configuration. \begin{table*}
\begin{center}
      \caption[]{Mass ($M$), compactness ratio $u(\equiv M/R)$ and fractional moment of inertia, $Q(\equiv I_{\rm core}/I_{\rm total)}$ for the  models discussed in the text. The various parameters are obtained  by  assigning a fiduciary value of the energy-density at the core-envelope boundary, $E_b  = 2.7 \times 10^{14}$ g\, cm$^{-3}$ for different values of   the ratio of pressure to energy-density, $(P_0/E_0)$, at the centre. In abovementioned parameters, the superscript a, b, and c  represent the models with an envelope $\Gamma_1  = (4/3), (5/3)$, and 2.0 respectively. The quantity $u_h$  represents  the  compactness ratio of homogeneous density distribution for the corresponding value of $(P_0/E_0)$. The 
{\em minimum} ratio of  pressure  to  energy-density, $(P_b/E_b) \cong 4.694 \times 10^{-2}$ ,  at  the 
core-envelope boundary is obtained in such a manner that for  an  assigned  value  of 
$(P_0/E_0)$ the inequality, $u \leq
u_h$,  is always satisfied for {\em all} models corresponding to different values of $\Gamma_1$ in the envelope.  The 
slanted values correspond to the limiting case upto which the configuration 
remains pulsationally stable.}

\begin{tabular}{ccccccccccccc}

\hline
${(P_0 / E_0)}$ & $u_h$  & $(M^a/M_{\odot})$ & $Q^a$ & $u^a$ & $(M^b/M_{\odot})$ & $Q^b$ & $u^b$ & $(M^c/M_{\odot})$ & $Q^c$ & $u^c$ \\ 
 
\hline

0.11236 & 0.15394 & 1.87195  & 0.02739 & 0.07042 & 1.67909  &  0.09920 & 0.12154  & 1.61142 &  0.13475 & 0.13537 \\
0.12029 & 0.16116 & 1.93294 & 0.03739 & 0.07693  & 1.76262  & 0.12071  & 0.12843 &  1.70131 & 0.15969 & 0.14236 \\
0.12955 & 0.16918 & 2.00865 & 0.05089 & 0.08466  &  1.85989 & 0.14626 & 0.13630 &  1.80486 & 0.18860 & 0.15034 \\
0.14549 & 0.18205 & 2.14512 & 0.07777 & 0.09798  & 2.02462 & 0.19020   & 0.14940 &  1.97821 & 0.23645 & 0.16340 \\
0.15354 & 0.18814 & 2.21518 & 0.09261 & 0.10461 & 2.10582 & 0.21182 & 0.15572 & 2.06296 & 0.25950 & 0.16969 \\
0.20061 & 0.21911 & 2.61355 & 0.18555 & 0.14077  & 2.54549 & 0.32652   & 0.18919 &  2.51677 & 0.37684 & 0.20255 \\
0.23990 & 0.24008 & 2.91143 & 0.26028 & 0.16698  &  2.86158 & 0.40470  & 0.21268 &  2.83972 & 0.45366 & 0.22536 \\
0.27901 & 0.25763 & 3.17036 & 0.32671 & 0.18963  & 3.13199 & 0.46877  & 0.23274  & 3.11471 & 0.51517 & 0.24469 \\
0.34816 & 0.28259 & 3.54016 & 0.42326 & 0.22255  & 3.51403 &  0.55590 & 0.26156 &  3.50187 & 0.59803 & 0.27253 \\
0.44840 & 0.30929 & 3.90788 & 0.52531 & 0.25816  & 3.89112 &  0.64342 & 0.29256  &  3.88312 & 0.67967 & 0.30222 \\
0.48306 & 0.31666 & 3.99636 & 0.55283 & 0.26795  & 3.98169 & 0.66607  & 0.30094 &  3.97464 & 0.70070 & 0.31024 \\
0.51773 & 0.32332 & 4.06762 & 0.57715 & 0.27667  & 4.05467 & 0.68625  & 0.30848 &  4.04842 & 0.71944 & 0.31746 \\
0.63150 & 0.34115 & 4.19170 & 0.63925 & 0.29895  & 4.18272 & 0.73770  & 0.32778 &  4.17831 & 0.76712 & 0.33589 \\
0.66866 & 0.34593 & {\sl 4.19788} & {\sl 0.65474} & {\sl 0.30438}  & {\sl 4.18981} &  {\sl 0.75034} & {\sl 0.33242}  &  {\sl 4.18584} & {\sl 0.77916} & {\sl 0.34040} \\
0.68296 & 0.34766 & 4.19565 & 0.66008 & 0.30623  & 4.18789 & 0.75491  & 0.33404 & 4.18408 & 0.78336 & 0.34194 \\
0.69900 & 0.34952 & 4.19032 & 0.66592 & 0.30820 &  4.18289 & 0.75976  & 0.33573 & 4.17923 & 0.78778 & 0.34352 \\
0.72302 & 0.35220 & 4.17593 & 0.67364 & 0.31076 &  4.16896 & 0.76625  & 0.33793 & 4.16552 & 0.79403 & 0.34567 \\
0.74010 & 0.35401 & 4.16090 & 0.67868 & 0.31235 &  4.15423 & 0.77068 & 0.33934 & 4.15094 & 0.79815 & 0.34700 \\
0.77503 & 0.35751 & 4.11752 & 0.68772 & 0.31501 &  4.11141 & 0.77855  & 0.34164 & 4.10839 & 0.80543 & 0.34914 \\
0.80605 & 0.36041 & 4.06269 & 0.69398 & 0.31651 &  4.05700 & 0.78427  & 0.34294 & 4.05419 & 0.81111 & 0.35044 \\
0.83972 & 0.36336 & 3.98260 & 0.69887 & 0.31711 &  3.97730 & 0.78917  & 0.34347 & 3.97469 & 0.81587 & 0.35091 \\

\hline

\end{tabular}

\end{center}
\end{table*}

The various $Q$ values obtained in this manner are also shown in Table 1 (the corresponding values of the moment of inertia of the entire configuration, $I_{\rm total,45}$, are not shown in this table)  for an assigned value of matching density $E_b = 2.7 \times 10^{14}\,{\rm g\, cm}^{-3}$. If the minimum values of the observational constraints, $Q\simeq 0.7$, and $I_{\rm Crab,45} \simeq 3.04$ are imposed on these models, the model with a  $\Gamma_1 = (4/3)$ envelope is ruled out because it always gives stable models with $Q < 0.7$, however, the model with a $\Gamma_1 = 2$ envelope yields the value of matching density, $E_b = 7.0794 \times 10^{14}{\rm\,g\,cm}^{-3}$, which satisfies the inequalities $Q \geq 0.7$ and $I_{\rm Crab,45} \geq 3.04$ for stable sequences of NS models corresponding to an envelope with $\Gamma_1 = (5/3)$ and 2 respectively. For the last value of matching density, the models with $\Gamma_1 = 2$ and (5/3) yield the minimum masses $M \simeq 2.455 M_\odot,\,2.532 M_\odot$ and the surface redshifts $z_R \simeq 0.6232,\,0.6384$ respectively. The corresponding $Q$ and $I_{\rm Crab,45}$ values are obtained as $Q\simeq 0.7,\,I_{\rm Crab,45} \simeq 3.04$ and $Q\simeq 0.7,\,I_{\rm Crab,45} \simeq 3.281$ for $\Gamma_1 = 2$ and (5/3) cases respectively. However, for the observational constraint of the ``central'' weighted mean value $Q \approx 0.72$, and $I_{\rm Crab,45} > 3.04$, the models with $\Gamma_1 = 2$ and (5/3) yield the masses $M \simeq 2.500 M_\odot,\,2.564 M_\odot$ and the surface redshifts $z_R \simeq 0.6550,\,0.6719$ respectively. The corresponding $I_{\rm Crab,45}$ values are obtained as $I_{\rm Crab,45} \simeq 3.105$ and $I_{\rm Crab,45} \simeq 3.293$ for $\Gamma_1 = 2$ and (5/3) cases respectively. It follows from these results that the NS models with a $\Gamma_1 = (5/3)$ envelope yields somewhat higher values of mass and surface redshift for the Crab pulsar as compared to $\Gamma_1 = 2$ envelope model, if the observational constraints of $Q$ and $I_{\rm Crab,45}$ are imposed.

(4) Page 676, right column, the statement ``(assuming $Q \geq 0.7$, as in the case of the Crab pulsar)'' of section 4 is removed and the statement in the second last sentence of this section is replaced by the following statement: For the case of $\Gamma_1 = 2$ envelope model, we get from Table 1 (after setting the matching density, $E_b = 7.0794 \times 10^{14}{\rm\,g\,cm}^{-3}$, and recalculating the parameters) the central value of surface redshift $z_R \simeq 0.23$ for a mass $M \simeq 1.274 M_\odot$ with $Q \simeq 0.2595$. The
binding energy corresponding to this case is obtained as $3.047 \times 10^{53}$ ergs which is
capable of releasing $3.047 \times 10^{44}$ ergs of energy required for the latter burst.

(5) Page 677, in the second line of first paragraph of section 5, read 
 $M_{\rm max} \approx 4.2 M_\odot$ instead of $M_{\rm max} \simeq 4.1 M_\odot$. The second sentence of the same paragraph is removed. In the same paragraph read 
$\Gamma_1 = 2$ instead of $\Gamma_1 = (5/3)$.
In the first sentence of second paragraph of section 5, read $\Gamma_1 = 2$ in place of $\Gamma_1 = (5/3)$ and read 0.6232 in place of 0.223 for a minimum surface redshift. Rest of the second paragraph of section 5 is the same as already discussed in the second sentence and onwards of point (1) above. 

\end{document}